\documentclass[aps,pra,showpacs,twocolumn, longbibliography]{revtex4-1} 

\usepackage{graphicx}
\usepackage{dcolumn}
\usepackage{bm}
\usepackage{color}
\usepackage{amsmath}
\usepackage{amssymb}
\begin{document}

\title{Mid-infrared spectrally-uncorrelated biphotons generation from doped PPLN: a theoretical investigation}

\author{Bei Wei$^{1, a}$}
\author{Wu-Hao Cai$^{1, a}$}
\author{Chunling Ding$^{1}$}
\author{Guang-Wei Deng$^{2, 3}$}
\email{gwdeng@uestc.edu.cn}
\author{Ryosuke Shimizu$^{4}$}
\author{Qiang Zhou$^{2, 3}$}
\email{zhouqiang@uestc.edu.cn}
\author{Rui-Bo Jin$^{1, 5}$}
\email{jrbqyj@gmail.com}


\affiliation{$^{1}$Hubei Key Laboratory of Optical Information and Pattern Recognition, Wuhan Institute of Technology, Wuhan 430205, PR China}
\affiliation{$^{2}$Institute of Fundamental and Frontier Sciences and School of Optoelectronic Science and Engineering, University of Electronic Science and Technology of China, Chengdu 610054, China}
\affiliation{$^{3}$CAS Key Laboratory of Quantum Information, University of Science and Technology of China, Hefei 230026, China}
\affiliation{$^{4}$The University of Electro-Communications, 1-5-1 Chofugaoka, Chofu, Tokyo, Japan}
\affiliation{$^{5}$State Key Laboratory of Quantum Optics and Quantum Optics Devices, Shanxi University, Taiyuan, 030006, China}
\affiliation{$^{a}$These authors contributed equally to this work}

\date{\today }


\begin{abstract}
We theoretically investigate the preparation of mid-infrared (MIR) spectrally-uncorrelated biphotons from a spontaneous parametric down-conversion process using doped LN crystals, including  MgO doped LN, ZnO doped LN,  and In$_2$O$_3$ doped ZnLN with doping ratio from 0 to 7 mol\%.
The tilt angle of the phase-matching function and the corresponding poling period are calculated under type-II, type-I, and type-0 phase-matching conditions.
We also  calculate the thermal properties of the doped LN crystals and their performance in Hong-Ou-Mandel interference.
It is found that the doping ratio has a substantial impact on the group-velocity-matching (GVM) wavelengths.
Especially,  the GVM$_2$ wavelength of co-doped InZnLN crystal has a tunable range of 678.7 nm, which is much broader than the tunable range of less  than 100 nm achieved by the conventional method of adjusting the temperature.
It can be concluded that the doping ratio can be utilized as a degree of freedom to manipulate the biphoton state.
The spectrally uncorrelated biphotons can be used to prepare pure single-photon source and entangled photon source,
which may have promising applications for quantum-enhanced sensing, imaging, and communications at the MIR range.
\end{abstract}

\maketitle
\section{Introduction}
The mid-infrared (MIR) wavelength range (approximately 2-20 $\mu$m) is of great interest to a variety of scientific and technological applications in sensing, imaging, and communications \cite{Tournie2019book, Ebrahim-Zadeh2008book}.
Firstly, this range contains strong absorption bands of a variety of gases, leads to essential  applications in gas sensing \cite{Shamy2020} and  environmental monitoring \cite{Chen2020}.
Secondly, the MIR region covers the atmospheric transmission window between 3 $\mu$m and 5 $\mu$m.  This region with relatively high transparency is  beneficial  for atmospheric monitoring and free-space communications \cite{Bellei2016oe}.
Thirdly, room temperature objects emit light at MIR wavelengths, resulting in novel applications in infrared thermal imaging \cite{Tittl2015, Mancinelli2017}.
Fourthly, with the rapid development of optical fiber communications, there is a growing demand to expand the communication wavelengths into the MIR region to  increase the communication bandwidth \cite{Soref2015}.

All these practical applications mentioned above are based on strong light sources in the MIR region.
However, the  advantages of the MIR range have not been fully exploited using quantum technology.
In order to further improve the sensitivity in those applications, one promising approach is to utilize a single-photon source or entangled photon source in the MIR region.
Spontaneous parametric down-conversion (SPDC) is one of the widely used methods to prepare biphotons, which can be used to produce single-photon source and entangled photon source.
Previously, several theoretical and experimental works have been dedicated to the development of single or entangled photon source in MIR range from an SPDC process.
In 2016, Lee et al. calculated the extended phase-matching properties of  periodically poled potassium niobate (PPKN)  for single-photon generation at the MIR range \cite{Lee2016}.
In 2017, Mancinelli et al. performed coincidence measurements on twin photons  generated from periodically poled lithium niobate (PPLN)  at 3.1 $\mu$m \cite{Mancinelli2017}.
In 2018,  McCracken, et al. numerically investigated the mid-infrared single-photon generation in a range of novel nonlinear  materials, including PPLN, PPKTP, GaP, GaAs, CdSiP$_2$, and ZnGeP$_2$ \cite{McCracken2018}.
In 2019, Rosenfeld et al. experimentally prepared MIR photon pairs in a four-wave mixing process from a silicon-on-insulator waveguide  at around 2.1 $\mu$m \cite{Rosenfeld2019}.
In 2020 Kundys et al. numerically studied the reconfigurable MIR single-photon  sources based on functional ferroelectrics, i.e., PMN-0.38PT crystal at 5.6 $\mu$m \cite{Kundys2020}.
In the same year, Prabhakar et al. experimentally demonstrated the entangled photons generation and Hong-Ou-Mandel interference at 2.1 $\mu$m from PPLN crystals \cite{Prabhakar2020}.

Many biphoton sources have been demonstrated in the previous studies \cite{Lee2016, Mancinelli2017, McCracken2018, Kundys2020, Prabhakar2020}; however, from the experimental point of view, it is still lack of high-quality biphotons source at the MIR range, especially the spectrally uncorrelated biphotons\cite{Mosley2008PRL, Grice2001, URen2006, Graffitti2018PRA,  Laudenbach2016}.
The biphotons generated from an SPDC process are generally spectrally correlated due to the energy and momentum conservation laws.
However, it is necessary to utilize biphotons with no spectral correlations to obtain the heralded pure-state single photon for many quantum information processing (QIP) protocols, such as quantum computation \cite{Walmsley2005}, boson sampling \cite{Broome2013}, and quantum teleportation \cite{Valivarthi2016np},  measurement-device-independent quantum key distribution \cite{Zhou2017QST}.
In quantum sensing, a high-NOON state, which provides high accuracy for phase-sensitive measurements, should be prepared  from spectrally uncorrelated biphotons \cite{Nagata2007, Yabuno2012PRA}.
Therefore, spectrally uncorrelation of photon pairs from SPDC is indispensable  for the future development of quantum information at the MIR range.

In this work, we investigate the generation of spectrally uncorrelated biphotons  from PPLN crystals at MIR range.
PPLN has the merits of large nonlinear coefficient and wide transparency range (0.4 $\mu$m $\sim$ 5 $\mu$m) \cite{Nikogosyan2005, Liu2017,Yang2020Josab, Wang2020PPLN3D}.
Another unique advantage of PPLN  is that its intrinsical group-velocity-matched (GVM) wavelengths are in the MIR range, and the spectrally uncorrelated biphoton state can be engineered at the GVM wavelengths \cite{Mosley2008PRL, Jin2018PRAppl}.

In addition, we also consider the manipulation of the biphoton state by adding different dopants in PPLN.
Traditionally, PPLN is doped with different dopants to improve its properties.
For example, by doping Mg and Zr, the damage resistance can be improved from visible to ultraviolet region \cite{Kong2020};
by doping Zn, the electro-optical coefficients can be improved \cite{Abdi1999};
by doping Fe, the photorefractive properties can be improved \cite{Zhang1998JAP}.
Especially, doping rare-earth elements (e.g., Tm, Er, Dy, Tb, Gd, Pr) makes PPLN a right candidate for lasers and quantum memories \cite{Palatnikov2006, Dutta2020nanolett}.
It is natural to deduce that doping can also be utilized as a degree of freedom to manipulate the single photons state at the MIR range.
Therefore, we  study the spectrally uncorrelated biphotons generation from doped PPLN in this work. Specifically, we investigated the GVM wavelengths, the tilt angle, the poling period, the thermal properties and the HOM interferences of biphtons generated from three kinds of doped LN crystals, including  MgLN [MgO($x$ mol\%)LiNbO$_3$], ZnLN [ZnO($x$ mol\%)LiNbO$_3$], and InZnLN [In$_2$O$_3$($x$ mol\%)ZnO(5.5 mol\%)LiNbO$_3$, with  $x$ from 0 to 7].

\section{Theory }
The biphoton state $\vert\psi\rangle$ generated from SPDC can be expressed as \cite{Mosley2008NJP}
\begin{equation}\label{eq1}
\vert\psi\rangle=\int_0^\infty\int_0^\infty\,\mathrm{d}\omega_s\,\mathrm{d}\omega_if(\omega_s,\omega_i)\hat a_s^\dag(\omega_s)\hat a_i^\dag(\omega_i)\vert0\rangle\vert0\rangle,
\end{equation}
where $\omega$ is the angular frequency, $\hat a^\dag$ is the creation operator, and the subscripts $s$ and $i$ indicate the signal and idler photon.
The joint spectral amplitude (JSA)  $f(\omega_s,\omega_i)$ can be calculated as the product of the pump envelope function (PEF) $\alpha(\omega_s,\omega_i)$ and the phase-matching function (PMF) $\phi(\omega_s,\omega_i)$:
\begin{equation}\label{eq2}
f(\omega_s,\omega_i) = \alpha(\omega_s,\omega_i) \times \phi(\omega_s,\omega_i).
\end{equation}
A PEF with a Gaussian-distribution can be written as
\begin{equation}\label{eq20}
\alpha(\omega_s, \omega_i)=\exp[-\frac{1}{2}\left(\frac{\omega_s+\omega_i-\omega_{p_0}}{\sigma_p}\right)^2],
\end{equation}
where $\omega_{p_0}$ is the center angular frequency of the pump; $\sigma_p$ is the bandwidth of the pump, and the subscript $p$  indicates the pump photon.  The full-width at half-maximum (FWHM) is FWHM$_\omega$ = 2$\sqrt{\ln(2)} \sigma_p \approx 1.67\sigma_p$.

If we choose wavelengths as the variables, the PEF can be rewritten as
\begin{equation}\label{eq21}
\alpha(\lambda_s, \lambda_i)= \exp \left(  -\frac{1}{2} \left\{  \frac{{1/\lambda _s  + 1/\lambda _i  - 1/(\lambda _0 /2)}}{{\Delta \lambda /[(\lambda _0 /2)^2  - (\Delta \lambda /2)^2 ]}}  \right\}^2 \right),
\end{equation}
where  $\lambda _0 /2 $ is the central wavelength of the pump; The FWHM of the pump at intensity level is
FWHM$_\lambda$ = $ \frac{2\sqrt{\ln (2)} {\lambda _0}^2   \Delta \lambda    \left({\lambda _0}^2-\Delta \lambda ^2\right)}{{\lambda _0}^4+\Delta \lambda ^4-2 {\lambda _0}^2 \Delta \lambda ^2 [1+\ln (4)]} $.
For $\Delta \lambda << \lambda _0$, FWHM$_\lambda\approx 2\sqrt{\ln(2)} \Delta \lambda  \approx 1.67\Delta \lambda $.

The PMF function can be written as
\begin{equation}\label{eq4}
\phi(\omega_s,\omega_i)=\operatorname{Sinc}\left(\frac{\Delta kL}{2}\right),
\end{equation}
where $L$ is the length of the crystal,  $\Delta k =k_p-k_i-k_s\pm\frac{2\pi}{\Lambda}$ and  $k=\frac{2 \pi n (\lambda)}{\lambda}$ is the wave vector.
$\Lambda$ is the poling period and can be calculated by
\begin{equation}\label{eq4}
\Lambda=\frac{2\pi}{|k_p-k_i-k_s|}.
\end{equation}
The angle $\theta$ between the ridge direction of PMF and the horizontal axis is determined by \cite{Jin2013OE}
\begin{equation}\label{eq6}
\tan\theta=-\left( \frac{V_{g,p}^{-1}(\omega_p)-V_{g,s}^{-1}(\omega_s)}{V_{g,p}^{-1}(\omega_p)-V_{g,i}^{-1}(\omega_i)} \right),
\end{equation}
where  $V_{g,\mu}=\frac{d\omega}{dk_\mu(\omega)}=\frac{1}{k_\mu^\prime(\omega)},(\mu=p, s, i)$ is the group velocity.
The shape of the JSA is determined by the following three GVM conditions  \cite{Edamatsu2011, Jin2019PRAppl}. 
 \\
The GVM$_1$ condition ($\theta=$ 0$^ \circ$):
\begin{equation}\label{gvm1}
V_{g,p}^{-1}(\omega_p)=V_{g,s}^{-1}(\omega_s).
\end{equation}
The GVM$_2$ condition ($\theta=$ 90$^ \circ$):
\begin{equation}\label{gvm2}
V_{g,p}^{-1}(\omega_p)=V_{g,i}^{-1}(\omega_i).
\end{equation}
The GVM$_3$ condition ($\theta=$ 45$^ \circ$):
\begin{equation}\label{gvm3}
2V_{g,p}^{-1}(\omega_p)=V_{g,s}^{-1}(\omega_s)+V_{g,i}^{-1}(\omega_i).
\end{equation}
The purity can be calculated by performing Schmidt decomposition on $f(\omega_s,\omega_i)$:
\begin{equation}\label{eq10}
f(\omega_s,\omega_i) = \sum_jc_j\phi_j(\omega_s)\varphi(\omega_i),
\end{equation}
where $\phi_j(\omega_s)$ and $\varphi_j(\omega_i)$ are the two orthogonal basis vectors in the frequency domain, and $c_j$ is a set of non-negative real numbers that satisfy the normalization condition $ \sum_jc_j^2=1$.
The purity p is defined as:
\begin{equation}\label{eq12}
p=\sum_jc_j^4.
\end{equation}
GVM wavelengths are important parameters for nonlinear crystals.
The maximal purities under GVM$_1$, GVM$_2$, and GVM$_3$ conditions are around 0.97, 0.97, and 0.82, respectively \cite{Edamatsu2011}.

Note the three GVM conditions in Eqs.(\ref{gvm1}-\ref{gvm3}) are also called fully-asymmetric GVM or symmetric GVM conditions in Ref. \cite{Graffitti2018PRA}.
In fact, the spectrally-pure state can be prepared for any $\theta$ angle between 0 and 90 degrees, independently from the PDC-type (type-0, type-I, or type-II) and wavelength degeneracy (degenerated or non-degenerated).
Perfectly separable JSA can be achieved only with Gaussian pump and Gaussian PMF (via domain engineering techniques), as proven in Ref. \cite{Quesada2018PRA}.
However, the three GVM conditions listed in Eqs.(\ref{gvm1}-\ref{gvm3}) with the degenerated wavelengths are the most widely used cases in the experiments \cite{Jin2020BBO, Cai2020JOSAB, Duan2020JosaB}. Especially, the GVM$_3$ condition has a symmetric distribution along the anti-diagonal direction, i.e., $f(\omega_s,\omega_i)=f(\omega_i,\omega_s)$, which is the prerequisite condition for high-visibility HOM interferences. Therefore, we mainly focus on the type-II phase-matching condition in the calculation below.

%
%
\begin{figure}[tbp]
\centering\includegraphics[width=8cm]{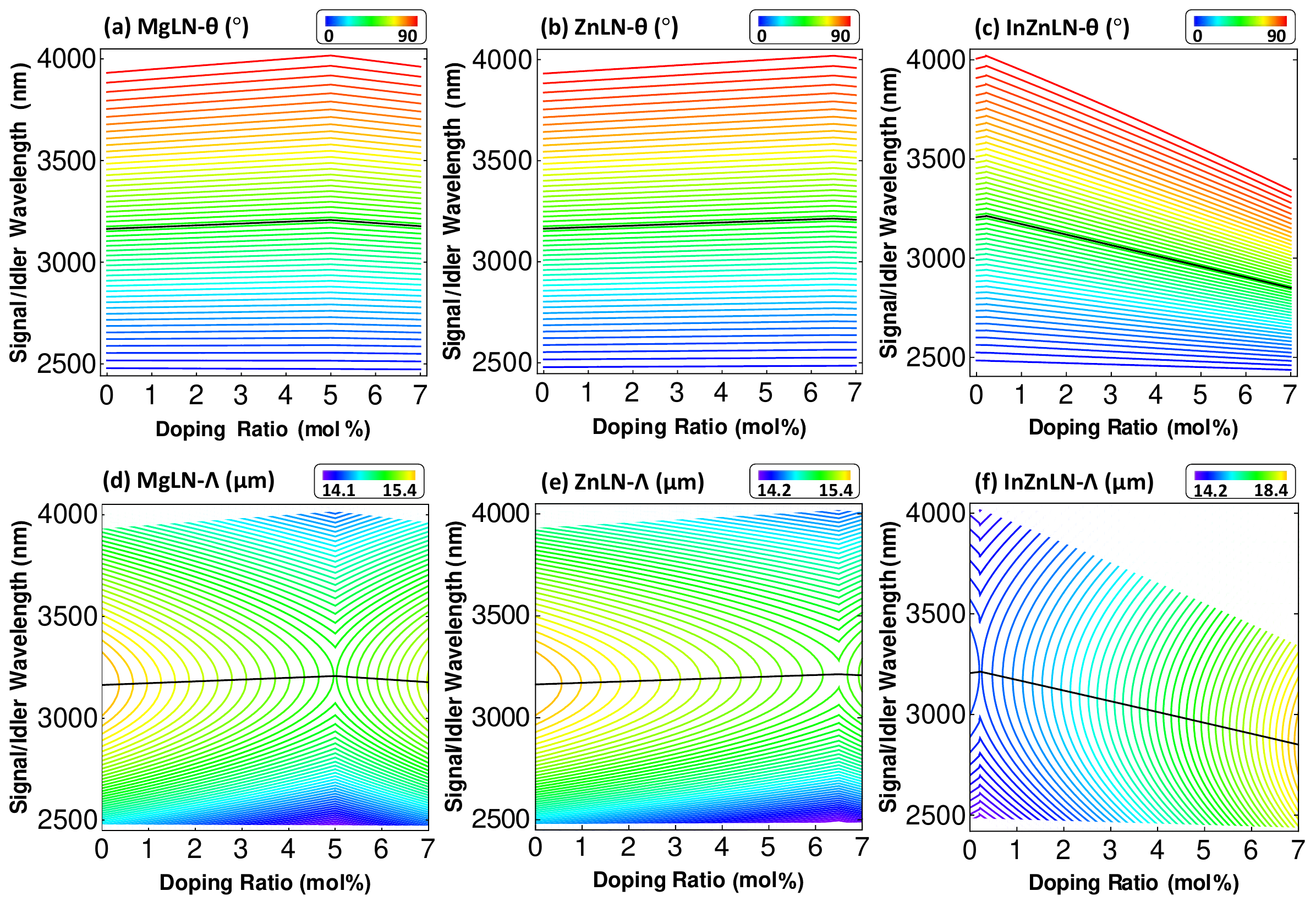}
\caption{(a-c): The tilt angle $\theta$  as a function of signal (idler) wavelength and doping ratio $x$  for MgLN [MgO($x$ mol\%):LiNbO3],  ZnLN [ZnO($x$ mol\%):LiNbO$_3$], and InZnLN [In$_2$O$_3$($x$ mol\%):ZnO(5.5 mol\%):LiNbO$_3$]. (d-f): The corresponding poling period. In these figures, the signal and the idler photons have  degenerated wavelengths under the \textbf{type-II} phase-matching condition. The  solid  black line indicates $\theta= 45^{\circ}$.
 } \label{typeII_dopratio}
\end{figure}
%

\begin{figure}[tbp]
\centering\includegraphics[width=8cm]{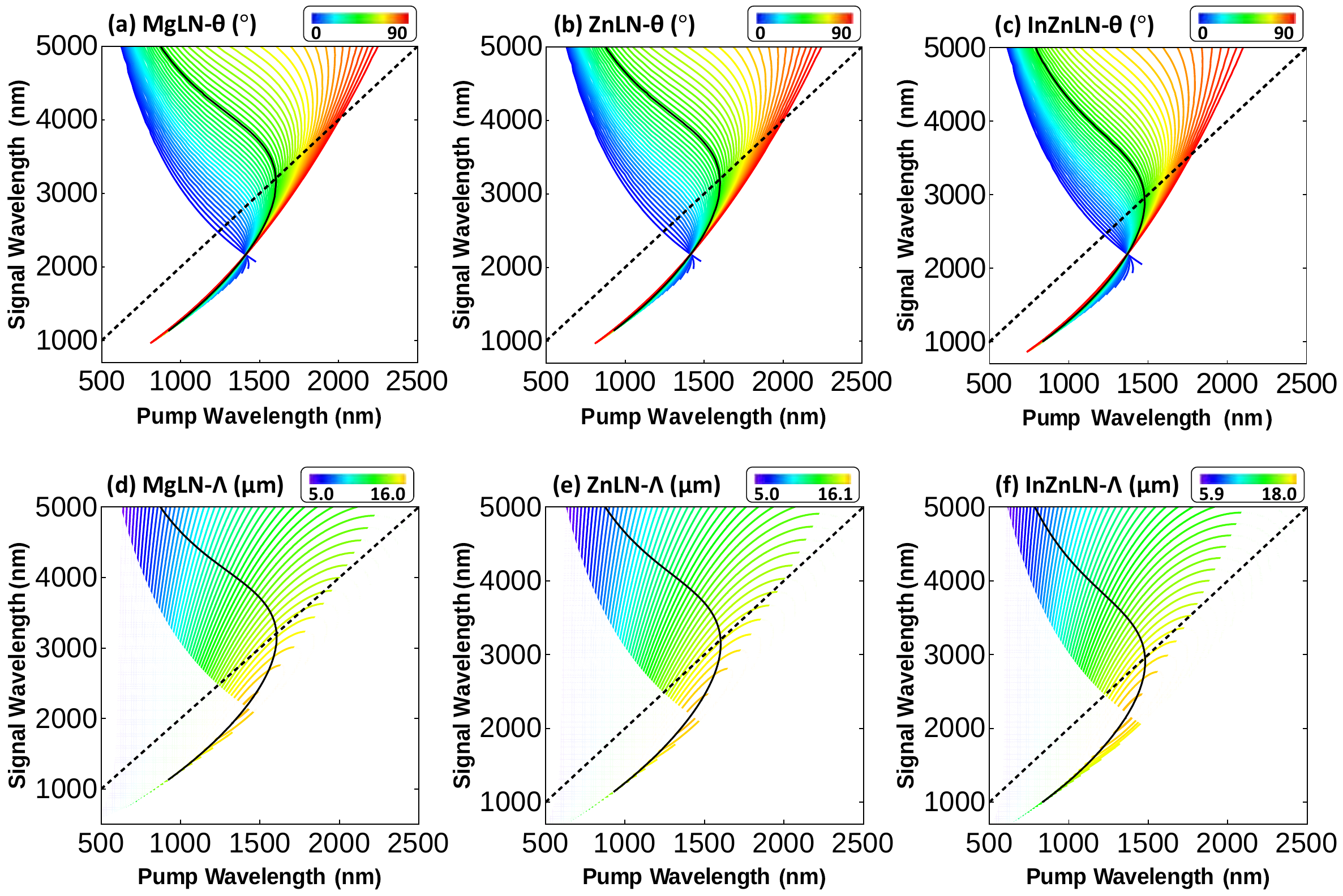}
\caption{(a-c): The tilt angle $\theta$  as functions of the signal wavelength and the pump wavelength for crystals with fixed doping ratio, i.e.,  MgLN [MgO($5$ mol\%):LiNbO3],  ZnLN [ZnO($5$ mol\%):LiNbO$_3$], and InZnLN [In$_2$O$_3$($5$ mol\%):ZnO(5.5 mol\%):LiNbO$_3$]. (d-f): The corresponding poling period. In these figures, the signal and the idler photons have the non-degenerated wavelengths under the \textbf{type-II} phase-matching condition. The  solid black line indicates $\theta= 45^{\circ}$, and the dashed black line indicates $\lambda_s = 2\lambda_p$.
 } \label{typeII_nondeg}
\end{figure}

\section{Calculation and simulation}
In this section, we first calculate the angle $\theta$ in the range of 0 and 90 degrees, and the corresponding poling period $\Lambda$ under type-II, type-I, and type-0 phase-matching conditions.
Then, we consider the thermal properties, and the HOM interferences for the doped PPLN crystals under the type-II phase-matching condition. The Sellmeier equations for these crystals are chosen from Refs. \cite{Schlarb1994, Schlarb1995,Schlarb1996}, where the Sellmeier equations are parameterized by the dopant concentration.
The PPLN crystals are negative uniaxial crystals $(n_o > n_e)$, where $n_{o(e)}$ is the refractive index of the ordinary (extraordinary) ray.

\subsection{Angle $\theta$ and poling period $\Lambda$ for type-II phase-matching condition}

In the calculation below, we  consider type-II phase-matched  ($o \to o + e$) SPDC in a collinear configuration with wavelength degenerated  $(2\lambda_p=\lambda_s=\lambda_i)$ or non-degenerated ($\lambda_s \neq \lambda_i)$.
In the type-II phase-matching condition, we set the pump and the signal as  o-ray,  and the idler as e-ray.
Figure\,\ref{typeII_dopratio}(a-c) shows the tilt angle $\theta$ as a function of the doping ratio ($x$ mol \%) and signal/idler (degenerated) wavelength.
For MgLN in Fig.\,\ref{typeII_dopratio}(a), on the line of  $\theta = 0^{\circ}$, the GVM$_1$ wavelength increased monotonically, i.e., from  2478.6 nm at $x=0$ to 2472.7 nm at $x=7$.
For $\theta = 90^{\circ}$, the   GVM$_2$ wavelength increased non-monotonically from 3931.4 nm ($x=0$) to the maximal value of 4016.6 nm ($x=5$), and then decreased to 3961.6 nm ($x=7$).
For $\theta = 45^{\circ}$, the   GVM$_3$ wavelength also increased non-monotonically, i.e., increased from 3163.6 nm ($x=0$) to 3207.6  nm ($x=5$), and then decreased to 3177.1 nm ($x=7$).

For  ZnLN in Fig.\,\ref{typeII_dopratio}(b), on the line of  $\theta = 0^{\circ}$, the GVM$_1$ wavelength increased monotonically from 2478.6 nm at $x=0$ to 2485.8 nm at $x=7$.
On the line of  $\theta = 90^{\circ}$, the GVM$_2$ wavelength start from 3931.4 nm at $x=0$, and reach a maximal value of 4019.3 nm at $x=6.5$, and then decreased to 4009.0 nm at $x=7$.
On the line of $\theta = 45^{\circ}$, the GVM$_3$ wavelength is 3163.6 nm, 3213.6 nm, and 3208.2 nm, at $x=0$, 6.5, and 7, respectively.

\begin{table}[tbh]
\centering
\begin{tabular}{cccc}
\hline \hline
Crystal Name                                               &MgLN        &ZnLN        &InZnLN   \\
\hline
$\lambda_\textrm{GVM1}$  at      0 mol\%  (nm)             &2478.6     &2478.6          &2484.3  \\
$\lambda_\textrm{GVM1}$  at      7 mol\%  (nm)             &2472.7     &2485.8          &2436.9          \\
$\Delta\lambda_\textrm{GVM1}$ (nm)                         &5.9     &-7.2        &47.4          \\
 \hline
$\lambda_\textrm{GVM2}$  at      0 mol\%  (nm)              &3931.4   &3931.4           &4006.0       \\
$\lambda_\textrm{GVM2}$  at      7 mol\%  (nm)              &3961.6   &4009.0           &3342.9             \\
Maximum of $\lambda_\textrm{GVM2}$                          &4016.6   &4019.3           &4021.6             \\
$\Delta\lambda_\textrm{GVM2}$ (nm)                          &-85.2      &-87.9          &678.7          \\

 \hline
$\lambda_\textrm{GVM3}$  at      0 mol\%  (nm)              &3163.6   &3163.6           &3205.9        \\
$\lambda_\textrm{GVM3}$  at      7 mol\%  (nm)              &3177.1   &3208.2           &2850.4       \\
Maximum of $\lambda_\textrm{GVM3}$                          &3207.6   &3213.6           &3213.9         \\
$\Delta\lambda_\textrm{GVM3}$ (nm)                          &-44.0    &-50.0            &363.5        \\

\hline
\hline
\end{tabular}
\caption{\label{table:dop} Summary of the GVM wavelengths in  Fig.\,\ref{typeII_dopratio} at different doping ratios. The maximal $\lambda_\textrm{GVM}$ is achieved at the doping ratio of 5 mol\% for MgLN, 6.5 mol\% for ZnLN, and 0.2 mol\% for InZnLN. $\Delta\lambda$ is the maximal difference of the GVM wavelengths.}
\end{table}

For InZnLN in Fig.\,\ref{typeII_dopratio}(c), the GVM$_1$ ($\theta = 0$) wavelength decreased monotonically from 2484.3 nm at $x=0$ to 2436.9 nm at $x=7$.
The GVM$_2$ ($\theta = 90^{\circ}$) wavelength increased non-monotonically: start from 4006.0 nm at $x=0$  and reach a maximal value of 4021.6 nm at  $x=0.2$, and then decreased to 3342.9 nm at $x=7$.
The GVM$_3$ ($\theta = 45^{\circ}$) wavelength also increased non-monotonically, the wavelengths are 3205.9 nm, 3213.9 nm, and 2850.4 nm at $x=0$, 0.2, and 7, respectively.
Figure\,\ref{typeII_dopratio}(d-f)  shows the poling period for each crystal. $\Lambda$  is 14.1 $\mu$m to 15.4 $\mu$m for MgLN,   14.2 $\mu$m to 15.4 $\mu$m for ZnLN, and 14.2 $\mu$m to 18.4 $\mu$m for InZnLN.
For comparison, we also list the wavelength of GVM$_1$, GVM$_2$, and GVM$_3$ at different doping ratios in Tab.\,\ref{table:dop}.
It can be discovered  in Tab.\,\ref{table:dop} that the co-doped crystal, InZnLN,  has a large tunable wavelength range of 678.7 nm. This feature is essential for quantum state engineering at the MIR wavelength range.

Next, we consider the wavelength non-degenerated cases for a fixed doping ratio, i.e., MgO($5$ mol\%):LiNbO3, ZnO($5$ mol\%): LiNbO$_3$, and In$_2$O$_3$($5$ mol\%): ZnO(5.5 mol\%): LiNbO$_3$.
Figure\,\ref{typeII_nondeg}(a-c) shows the angle $\theta$ as a function of the pump and the signal wavelengths for MgLN, ZnLN, InZnLN, with the same doping ratio of 5 mol\%.
On the line of $\theta = 0^{\circ}$ ($\theta = 90^{\circ}$), the signal wavelength decreases (increases) monotonically with the increase of the pump wavelength.
On the line of $\theta = 45^{\circ}$, the signal wavelength changes non-monotonically.
Figure\,\ref{typeII_nondeg}(d-f)  shows the corresponding poling period  $\Lambda$, ranging from 5.0 $\mu$m to 18.0 $\mu$m for the three crystals.

\begin{figure}[tbp]
\centering\includegraphics[width=8cm]{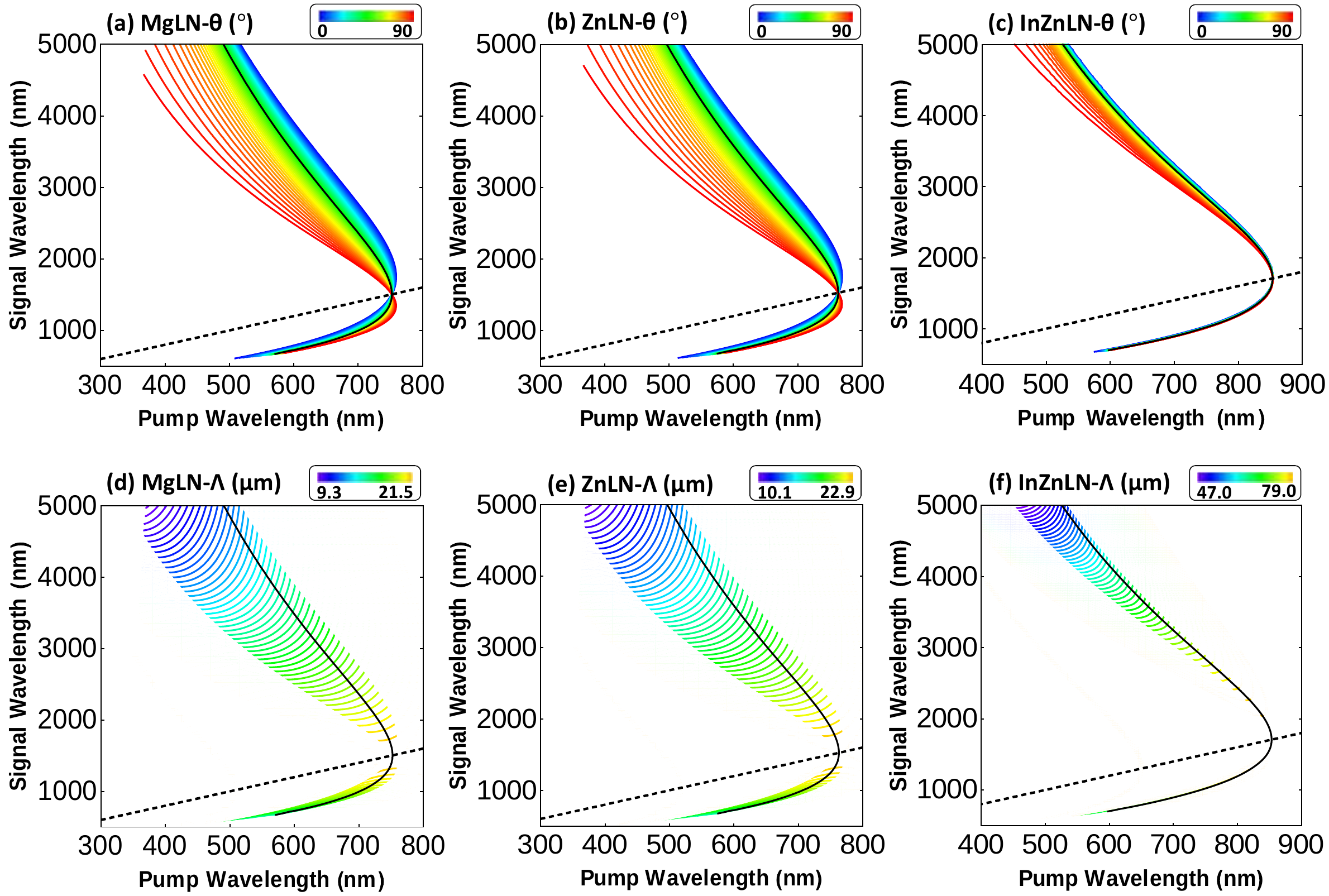}
\caption{(a-c): The tilt angle $\theta$  as a function of the signal wavelength and the pump wavelength for crystals with fixed doping ratio, i.e.,  MgLN [MgO($5$ mol\%):LiNbO3],  ZnLN [ZnO($5$ mol\%):LiNbO$_3$], and InZnLN [In$_2$O$_3$($5$ mol\%):ZnO(5.5 mol\%):LiNbO$_3$]. (d-f): The corresponding poling period. In these figures, the signal and the idler photons have the non-degenerated wavelengths under \textbf{type-I} phase-matching condition.  The  solid black line indicates $\theta= 45^{\circ}$, and the dashed black line indicates $\lambda_s = 2\lambda_p$.
 } \label{typeI_nondeg}
\end{figure}
%
%
%
\begin{figure}[tbp]
\centering\includegraphics[width=8cm]{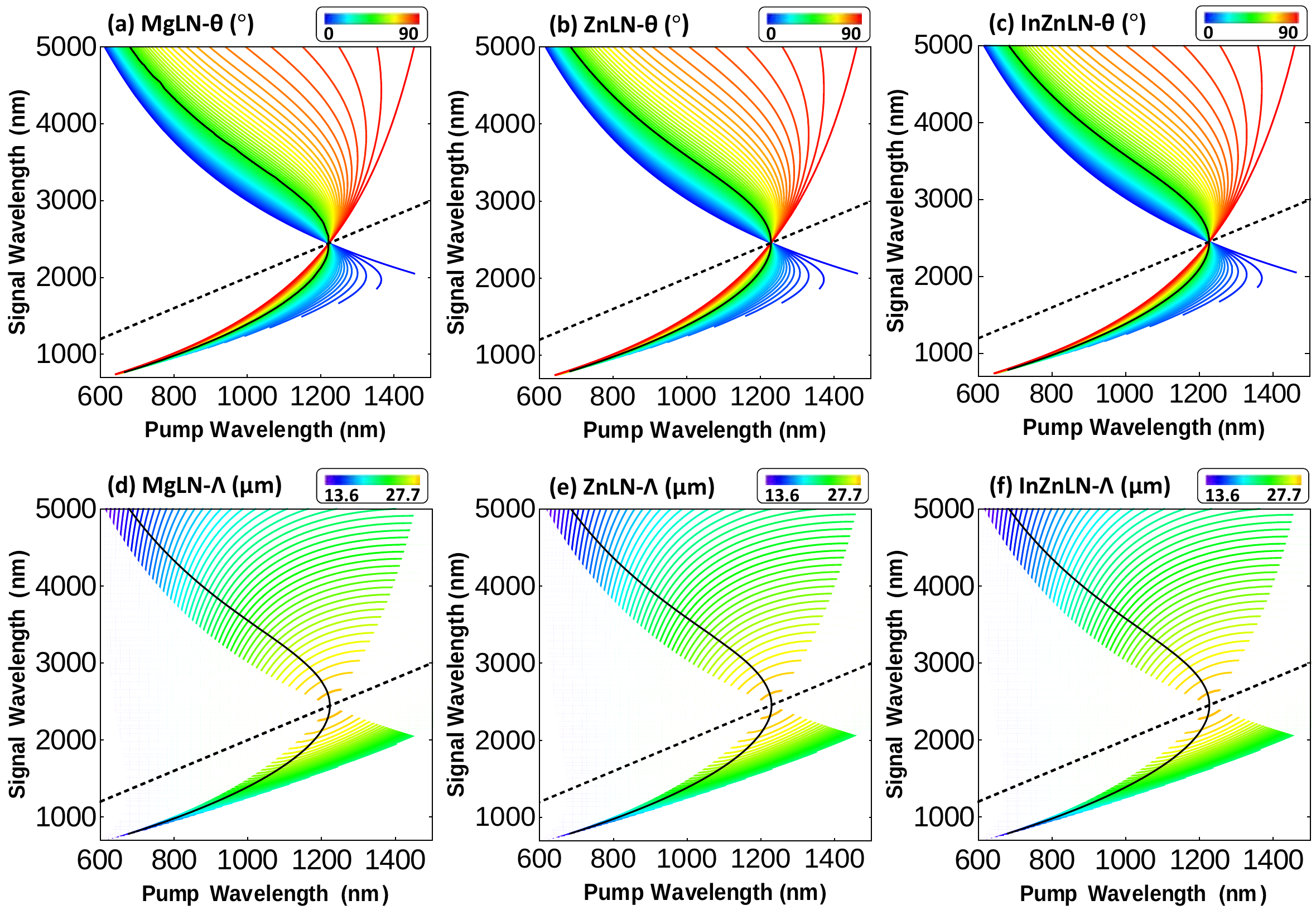}
\caption{(a-c): The tilt angle $\theta$  as functions of the signal wavelength and the pump wavelength for crystals with fixed doping ratio, i.e.,  MgLN [MgO($5$ mol\%):LiNbO3],  ZnLN [ZnO($5$ mol\%):LiNbO$_3$], and InZnLN [In$_2$O$_3$($5$ mol\%):ZnO(5.5 mol\%):LiNbO$_3$]. (d-f): The corresponding poling period. In these figures, the signal and the idler photons have the non-degenerated wavelengths under \textbf{type-0} phase-matching condition.  The  solid black line indicates $\theta= 45^{\circ}$, and the dashed black line indicates $\lambda_s = 2\lambda_p$.
 } \label{type0_nondeg}
\end{figure}

\subsection{Angle $\theta$ and poling period $\Lambda$ for type-I and type-0 phase-matching conditions}
It is also possible to prepare spectrally pure states under the type I and type-0 phase-matching conditions, as long as the angle $\theta$ is between 0$^{\circ}$  and 90$^{\circ}$.
Type-0 and type-I PDC in PPLN are routinely used in many quantum optics experiments, e.g., in biphoton states generation and frequency conversion \cite{Brecht2015PRX, Ansari2018Optica}.
Figure\,\ref{typeI_nondeg}(a-c) shows the angle $\theta$ as a function of the pump and the signal wavelengths under type-I ($e \to o + o$) phase-matching condition.
The pump wavelength is in the range of 300-900 nm and the signal is in the range of 500-5000 nm.
Figure\,\ref{typeI_nondeg}(d-f) shows the corresponding poling period  $\Lambda$,  ranging from 9.3 $\mu$m to 79.0 $\mu$m.
For wavelengths near the degenerated case, the poling period is larger than the case of far from the degenerated condition.
The  $\Lambda$ of InZnLN is larger than the values of MgLN and ZnLN.

Figure\,\ref{type0_nondeg}(a-c) shows the case for type-0 ($e \to e + e$) phase-matching condition.
The pump wavelength is in the range of 600-1500 nm,  and the signal is in the range of 700-5000 nm.
The corresponding poling period  $\Lambda$ is between 13.6 $\mu$m to 27.7 $\mu$m, as shown in Fig.\,\ref{type0_nondeg}(d-f).
The three crystals have a similar performance under the type-0 phase matching-condition.
\begin{figure}[tbp]
\centering\includegraphics[width=8cm]{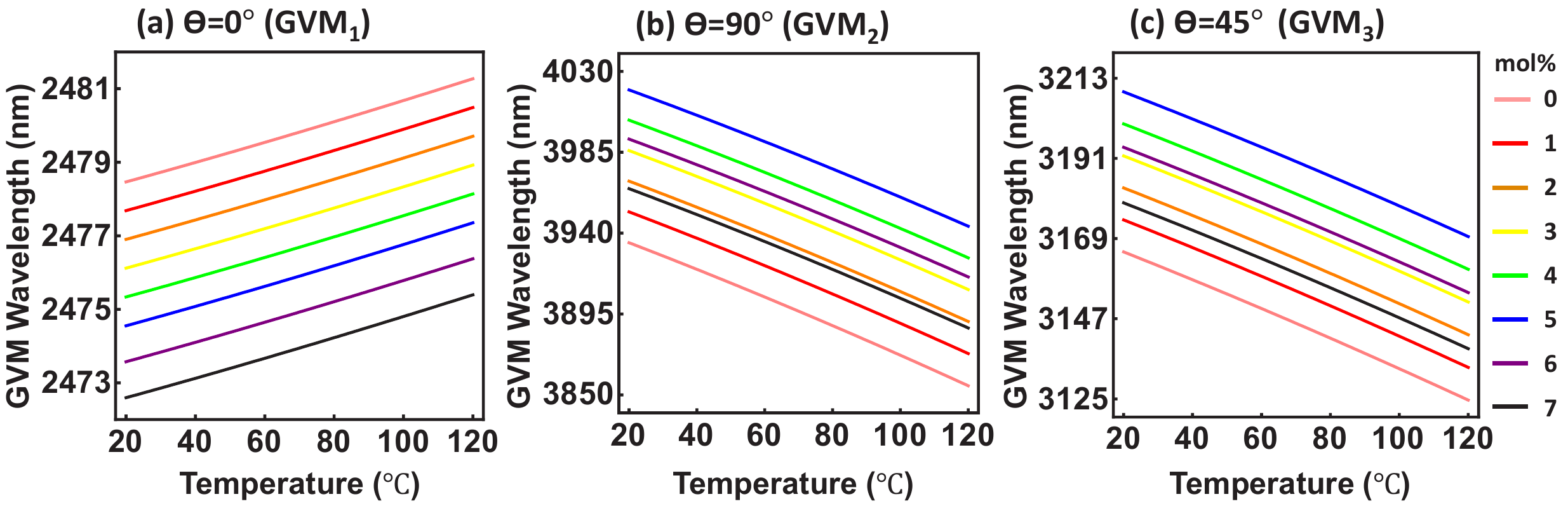}
\caption{Three GVM wavelengths of MgLN [MgO($x$ mol\%):LiNbO3] with different doping ratio $x$ (mol\%) at different temperatures.
 } \label{Tempr_LN}
\end{figure}
%
%
%
\begin{figure}[tbp]
\centering\includegraphics[width=8cm]{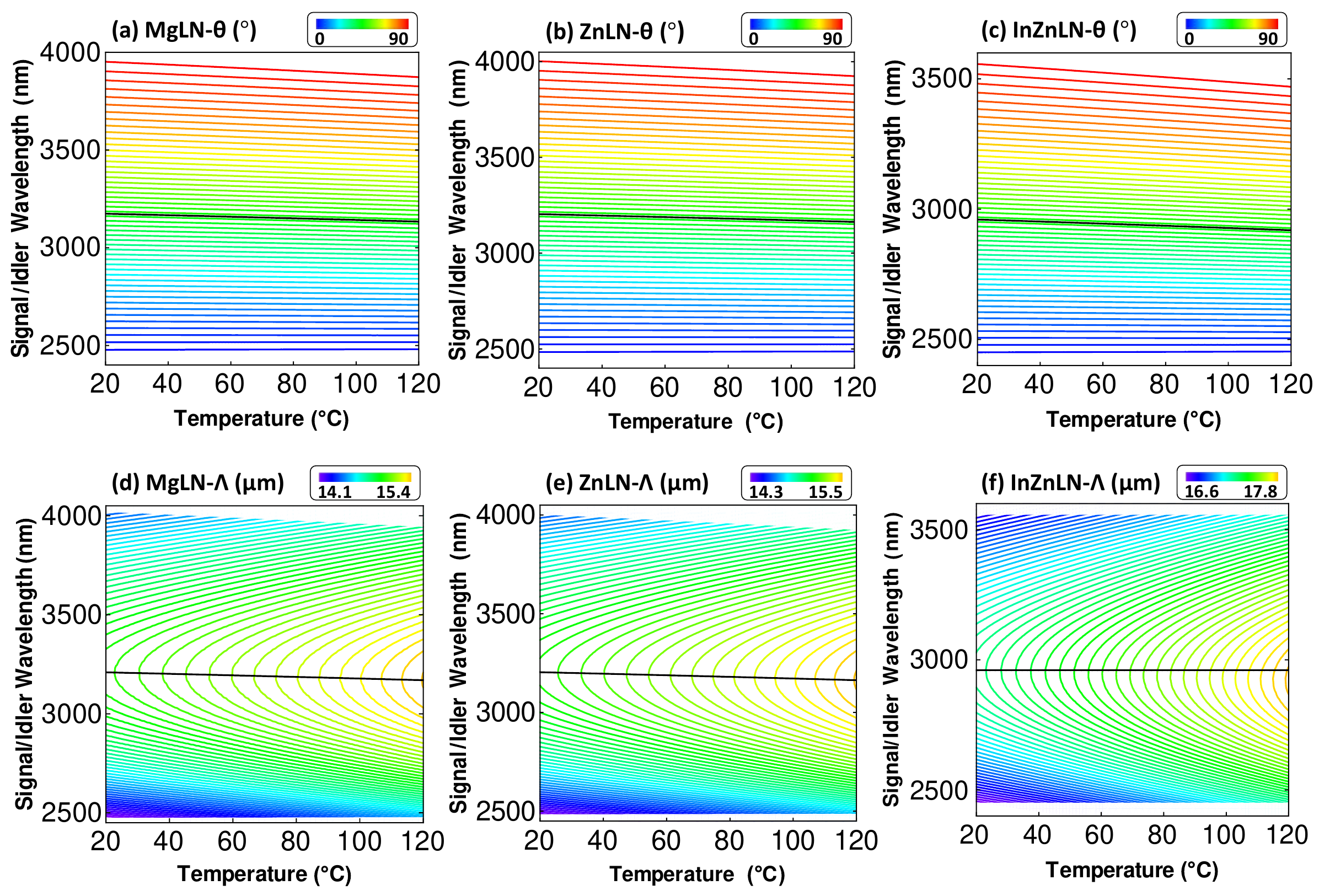}
\caption{(a-c): The tilt angle $\theta$  as functions of signal (idler) wavelength and temperature for crystals with fixed doping ratio, i.e.,  MgLN [MgO($5$ mol\%):LiNbO3],  ZnLN [ZnO($5$ mol\%):LiNbO$_3$], and InZnLN [In$_2$O$_3$($5$ mol\%):ZnO(5.5 mol\%):LiNbO$_3$]. (d-f): The corresponding poling period. In these figures, the signal and the idler photons have the degenerated wavelengths under \textbf{type-II} phase-matching condition. The  solid black line indicates $\theta= 45^{\circ}$, and the dashed black line indicates $\lambda_s = 2\lambda_p$. } \label{typeII_temperature}
\end{figure}
%
%
\begin{table}[tbh]
\centering
\begin{tabular}{cccc}
\hline \hline
Crystal Name                               &   MgLN                &ZnLN                   &InZnLN   \\
\hline
$\lambda_\textrm{GVM1}$  at      20 $\,^{\circ}\mathrm{C}$ (nm)           &2474.6     &2483.7           &2450.7  \\
$\Delta\lambda_\textrm{GVM1}$   20-120 $\,^{\circ}\mathrm{C}$  (nm)      &-2.8    &-2.8      &-2.8          \\
 \hline
$\lambda_\textrm{GVM2}$  at      20$\,^{\circ}\mathrm{C}$  (nm)           &4019.7   &4002.5     &3557.3       \\
$\Delta \lambda_\textrm{GVM2}$   20-120$\,^{\circ}\mathrm{C}$ (nm)        &75.9     &78.1       &87.3           \\
 \hline
$\lambda_\textrm{GVM3}$  at      20$\,^{\circ}\mathrm{C}$ (nm)           &3209.3   &3203.8         &2960.2        \\
$\Delta \lambda_\textrm{GVM3}$    20-120$\,^{\circ}\mathrm{C}$  (nm)      &39.7    &40.5       &41.1        \\
\hline
$\Lambda_\textrm{GVM1}$  at      20$\,^{\circ}\mathrm{C}$ ($\mu$m)          & 14.1&    14.2&        16.6        \\
$\Lambda_\textrm{GVM2}$  at      20$\,^{\circ}\mathrm{C}$ ($\mu$m)           &14.4&    14.5&        16.7        \\
$\Lambda_\textrm{GVM3}$  at      20$\,^{\circ}\mathrm{C}$ ($\mu$m)          & 15.0&    15.1&        17.2        \\
\hline
References                         &\cite{Schlarb1994}    &\cite{Schlarb1995}                   &\cite{Schlarb1996}  \\
\hline
\hline
\end{tabular}
\caption{\label{table2} Comparison of the GVM wavelengths $\lambda_\textrm{GVM1(2, 3)}$ at 20 $\,^{\circ}\mathrm{C}$, the wavelength difference $\Delta \lambda_\textrm{GVM1(2, 3)}$ between 20$\,^{\circ}\mathrm{C}$ and 120$\,^{\circ}\mathrm{C}$,  and the poling period $\Lambda_\textrm{GVM1(2, 3)}$ under three GVM conditions for three doped LN crystals: Mg(5 mol\%)LN, Zn(5 mol\%)LN and In(5 mol\%)Zn(5.5 mol\%)LN. }
\end{table}

\subsection{Thermal properties}
In this subsection, we investigate the thermal properties  of the doped PPLN crystals. Temperature is a crucial parameter for quasi-phase-matched (QPM) crystals, since the phase-matching conditions in the QPM crystals are mainly controlled by temperature in the experiments.
Using the temperature-dependent Sellmeier equations in Refs. \cite{Schlarb1994, Schlarb1995,Schlarb1996}, we can calculate the GVM wavelength as a function of the temperature.
We first consider the case of MgLN with different doping ratios, and then we compare MgLN with ZnLN and InZnLN at a fixed doping ratio.
Figure\,\ref{Tempr_LN}(a) shows the case of GVM$_1$ wavelength, which increases linearly with the increase of temperature.
When the temperature increases from 20$^{\circ}C$ to 120$^{\circ}C$, the GVM wavelength increases by about 2.8 nm for all the doping ratio from 0 to 7 mol\%.
Figure\,\ref{Tempr_LN}(b) and (c) show the cases of  GVM$_2$  and  GVM$_3$  wavelengths, which decrease by about 78 nm and 40 nm  respectively when the temperature increases from 20$^{\circ}C$ to 120$^{\circ}C$.
By comparing Figs.\,\ref{Tempr_LN}(a), (b),  and (c), it can be concluded that the GVM$_2$ wavelength is the most sensitive one to the temperature increase, i.e., decreases by about 78  nm. In contrast, GVM$_1$ is not so sensitive, i.e., only increase by 2.8 nm.

Next, we consider the thermal properties of other doped LN crystals. We compare the case of Mg(5 mol\%)LN, Zn(5 mol\%)LN, and In(5 mol\%)Zn(5.5  mol\%)LN in Fig.\,\ref{typeII_temperature} and Tab.\,\ref{table2}.
The most apparent feature in Fig.\,\ref{typeII_temperature} and Tab.\,\ref{table2} is that the signal/idler wavelength changes linearly. The wavelength difference between 20 $\,^{\circ}\mathrm{C}$ and 120 $\,^{\circ}\mathrm{C}$,   $\Delta \lambda_\textrm{GVM}$, is almost the same for three different crystals, i.e., $\Delta\lambda_\textrm{GVM1}$,  $\Delta\lambda_\textrm{GVM2}$,  and $\Delta\lambda_\textrm{GVM3}$  are about 2.8 nm, 78 nm, and 40 nm, respectively.
This phenomenon implies that the thermal-related GVM wavelength is not dominated by the doped elements, but by the main component, LiNbO$_3$.
The calculated the poling period of  the three crystals at 20 $\,^{\circ}\mathrm{C}$ are in the range of 14-15 $\mu$m, which can be realized with the current technology.
Note that the thermal expansion of the LN crystals are not considered in the calculation, since such data are not available in the references.

\subsection{Hong-Ou-Mandel interference}
\begin{figure}[tbp]
\centering\includegraphics[width=8cm]{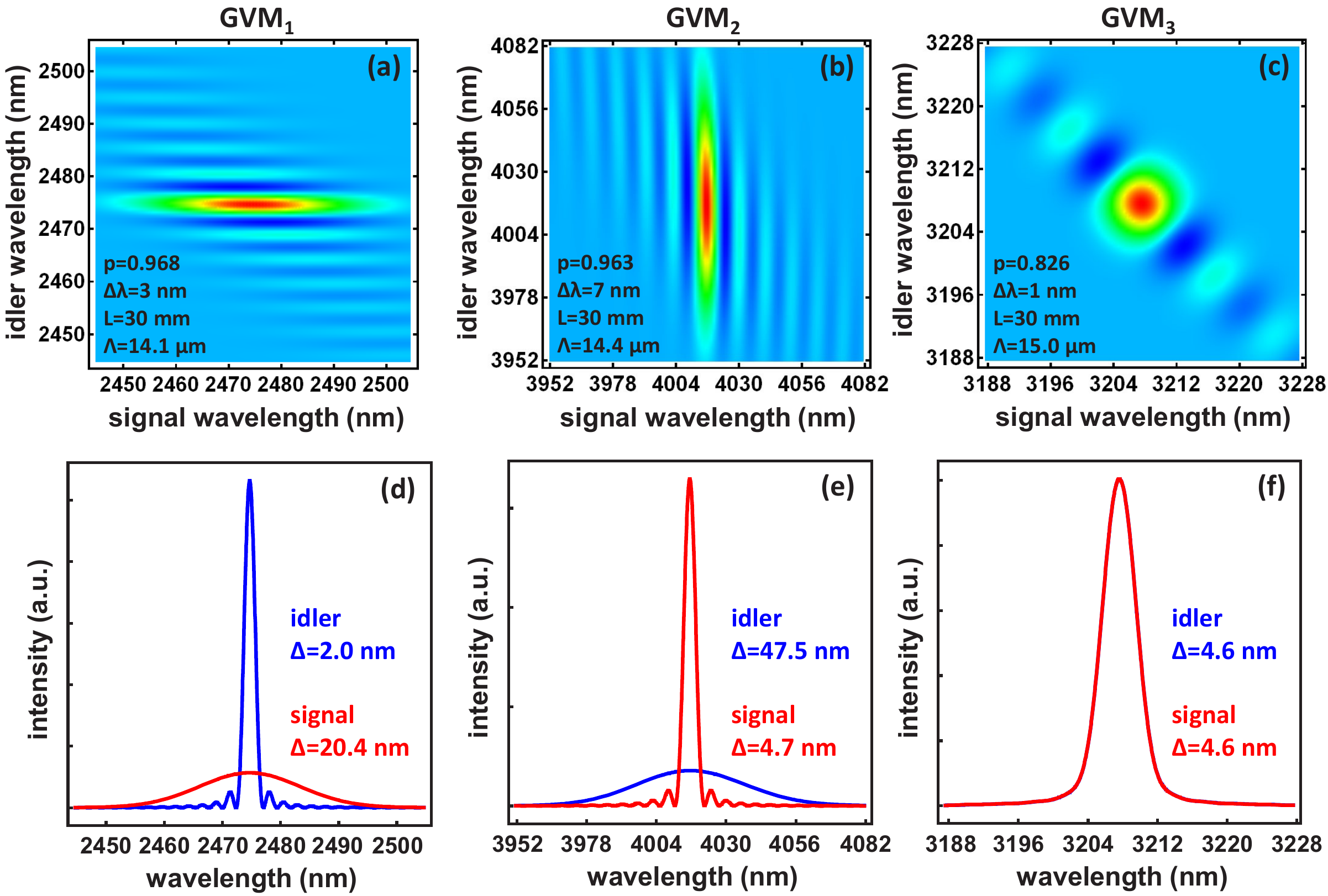}
\caption{The JSA and spectra of biphotons generated from 5 mol\% MgO doped PPLN under GVM$_1$, GVM$_2$, and GVM$_3$ conditions. The purity p, the pump bandwidth $\Delta \lambda$, the crystal length L, the poling period $\Lambda$, and FWHM values ($\Delta$) are listed in the figure. Note the blue curve and red curve are overlapped in (f).
 } \label{JSA}
\end{figure}
\begin{figure}[tbp]
\centering\includegraphics[width=8cm]{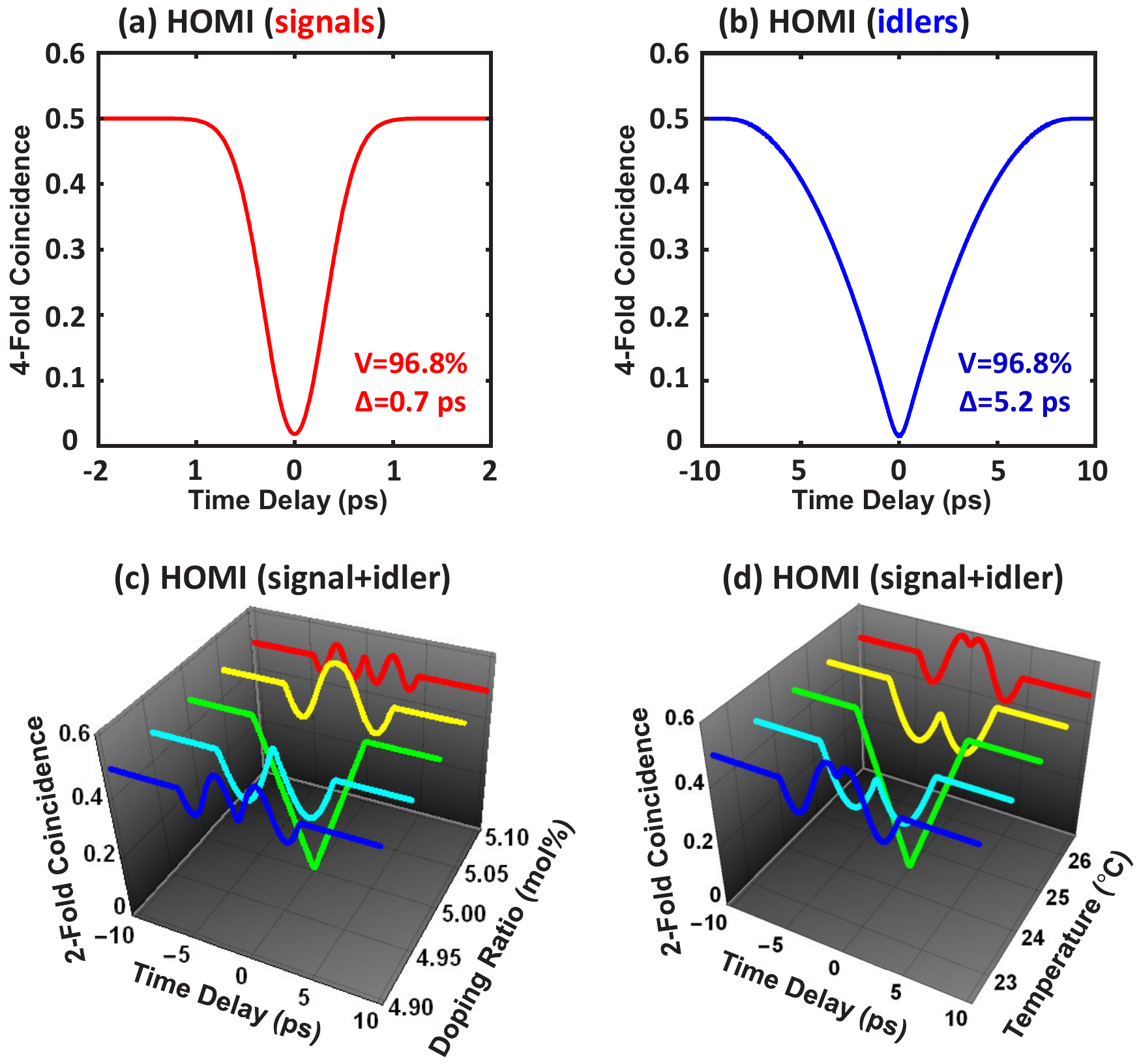}
\caption{ (a) and (b) are the HOM interference patterns using two heralded signal or idler photons from two independent PPMgLN crystals, with the  JSA shown in Fig.\,\ref{JSA}(a). (c) and (d) are the HOM interference patterns at different doping ratios and different temperatures, using the signal and idler photons from the same PPMgLN crystal, with the initial JSA shown in Fig.\,\ref{JSA}(c). The visibility (V) and FWHM of the pattern ($\Delta$) are listed in each figure.
 } \label{HOMI}
\end{figure}

In this subsection, we consider the  possible future applications of the spectrally-uncorrelated biphotons in Hong-Ou-Mandel interference \cite{Hong1987, Branczyk2017}, which is the foundation for many QIP applications such as quantum computation \cite{Walmsley2005}, boson sampling \cite{Broome2013}, and quantum teleportation \cite{Valivarthi2016np}.
Figure\,\ref{JSA} shows the JSA of the biphotons generated from 5 mol\% MgO doped PPLN under GVM$_1$, GVM$_2$, and GVM$_3$ conditions.
The spectral ranges for Figs.\,\ref{JSA}(a-c) are 60 nm, 130 nm, and 40 nm, respectively. The dimension of each JSA matrix is 200$\times$200.  Note the photon bandwidths shown in Fig.\,\ref{JSA} are just examples. In fact, the bandwidth is strongly dependent on the pulse duration and the crystal length, and the same photon purity can be achieved at different combinations of pump pulse and crystal length.
The biphotons under GVM$_1$ and GVM$_2$ conditions (see Fig.\,\ref{JSA}(a, b)) can be used for HOM interference between  two independent sources, with a typical experimental setup shown in Refs. \cite{Mosley2008PRL, Jin2013PRA}. In the experiment, two signals $s_{1}$ and $s_{2}$ are sent to a beamsplitter for interference and then detected by two single-photon detectors, while two idlers $i_{1}$ and $i_{2}$ are directly sent to single-photon detectors for heralding the signals.
The four-fold coincidence probability $P_4$ as a function of the delay time $\tau$ can be described by \cite{Ou2007, Mosley2008PRL, Jin2015OE}
\begin{equation}\label{eq:P4}
\begin{split}
P_4 (\tau )  = & \frac{1}{4}  \int_0^\infty \int_0^\infty \int_0^\infty \int_0^\infty d\omega _{s_1} d\omega _{s_2} d\omega _{i_1} d\omega _{i_2}  \\ & {\rm{|}}f_1 (\omega _{s_1} ,\omega _{i_1} )f_2 (\omega _{s_2} ,\omega _{i_2} )- \\ & f_1 (\omega _{s_2} ,\omega _{i_1} )f_2 (\omega _{s_1} ,\omega _{i_2} )e^{ - i(\omega _{s_2}  - \omega _{s_1} )\tau } {\rm{|}}^{\rm{2}},
\end{split}
\end{equation}
where $f_1$ and $f_2$ are the JSAs from the first and the second crystals.
Figure\,\ref{HOMI}(a, b) are the simulated HOM interference curves between two heralded signals or two heralded idlers from two independent MgLN sources, with the JSA shown in Fig.\,\ref{JSA}(a).
Without using any narrow bandpass filters, visibility can achieve 96.8\%.
It can be deduced from Eq.(\,\ref{eq:P4}) that the visibility, in this case, is determined by the purity (JSA separability) of the heralded signal photons \cite{Mosley2008PRL}.
The bandwidth of the HOM dip depends on the photon bandwidth, which is determined by the crystal length and the pulse duration. If we change the crystal length to be 50 mm, the corresponding HOM dips are 0.64 ps for two signals, and 8.4 ps for two idlers.

The biphotons under the GVM$_3$ condition (see Fig.\,\ref{JSA}(c)) can be used for HOM interference between the signal and the idler photons from the same SPDC source, with a typical setup shown in \cite{Hong1987, Prabhakar2020}. In the experiment,  the signal $s$ and the idler $i$ are sent to a beamsplitter for interference and then detected by two single-photon detectors.
The two-fold coincidence probability $ P_2(\tau )$ as a function of  $\tau$  is given by \cite{Grice1997, Ou2007, Jin2015OE}:
\begin{equation}\label{eq:P2}
\begin{split}
P_2(\tau)  =  & \frac{1}{4} \int\limits_0^\infty  \int\limits_0^\infty  d\omega _s  d\omega _i
                \left| {[f(\omega _s ,\omega _i ) - f(\omega _i ,\omega _s )e^{ - i(\omega _s  - \omega _i )\tau } ]} \right|^2.
\end{split}
\end{equation}
The green curve in Fig.\,\ref{HOMI}(c) is the simulated HOM interference pattern between a signal and an idler from the same Mg(5 mol\%)LN crystal at 24.5$^ \circ$C, with the JSA shown in Fig.\,\ref{HOMI}(c). Under this condition, the visibility is as high as 100\%.
By fixing the poling period and changing the doping ratio to 4.90 mol\%, 4.95 mol\%, 5.05 mol\%, and 5.10 mol\%, the interference patterns oscillate, as shown in Fig.\,\ref{HOMI}(c).
By changing the temperature to 22.5$^\circ$C, 23.5$^\circ$C,  25.5$^\circ$C and 26.5$^ \circ$C,  the interference patterns also oscillate, as shown in Fig.\,\ref{HOMI}(d).

\section{Discussions}
We notice that LN crystal has several different growing methods,  and each method has a different effect on the Sellmeier equations. The  MgLN, ZnLN, and InZnLN crystals investigated in the Calculation and Simulation section were grown using the Czochralski technique from a congruent melt (mole ratio Li/Nb  $\approx$ 0.942) \cite{Schlarb1994,Schlarb1995,Schlarb1996}.
As shown in Fig.\,\ref{typeII_dopratio}, 0 mol\% Zn doped LN (i.e., pure LN) has  the GVM$_1$,  GVM$_2$,  GVM$_3$ wavelengths of  2478.6 nm,  3931.4 nm , and 3163.6 nm, respectively.
In contrast, the pure LN grown from the stoichiometric melt (mole ratio Li/Nb  $\approx$ 1) \cite{Hobden1966} has the GVM$_1$,  GVM$_2$,  GVM$_3$ wavelengths of  2681.0 nm, 4653.6 nm, and 3595.8 nm, respectively.
Further, using the Sellmeier equations from Ref. \cite{Zelmon1997}, the pure LN grown from the congruent melt (mole ratio Li/Nb  $\approx$ 0.937) has the GVM$_1$,  GVM$_2$,  GVM$_3$ wavelengths of  2678.5 nm, 4361.7 nm, and 3499.1 nm, respectively.
These differences are mainly caused by different growing methods.
In addition, the LN crystal can also be doped with other elements, which also affect the Sellmeier equations. For example, NdMgLN crystal (grown by the Czochralski technique from a congruent melt in Ref. \cite{Kitaeva1998} has the GVM$_1$,  GVM$_2$,  GVM$_3$ wavelengths of 2615.3 nm, 4562.5  nm, and 3511.5 nm, respectively.
We also investigated ErMgLN and MgTiLN crystal \cite{Zhang2009,Yongmao1992}. However, they do not satisfy the  three kinds of GVM conditions.
In this calculation, the maximal doping ratio we considered is 7 mol\%. In fact, this ratio can still be improved, and we anticipate the wavelength tunability can be further improved at a larger doping ratio.

In this work, limited by the available Sellmeier equations, we only investigate the LN crystals doped with Mg, Zn, In, and Nd.
In the future, it is promising to explore more LN crystals doped with different chemical elements and with different doping ratios.
Another promising direction
 is investigating  the spectrally pure single-photon state generation from doped PPLN waveguide \cite{Cheng2019,Sun2019,Niu2020} or  doped PPLN film \cite{Ge2018}, which has the merits of higher nonlinear efficiency,  easier for integration  and microminiaturization.
In Fig.\,\ref{JSA}, the purities are 0.968, 0.963, and 0.826, respectively, and
this purity can be further improved to near 1 using the  custom poling technique \cite{Branczyk2011, Cui2019PRAppl}.

In the section of Calculation and Simulation, we only show the configuration of  $o \to o + e$ in the type-II phase-matching condition. In fact, we also investigate the configuration of  $e \to o + e$.   Although the GVM$_2$  wavelength under this configuration is in the telecom wavelength, the effective nonlinear coefficient, unfortunately,  is zero. The reason is analyzed in detail in the Appendix.

For future applications, spectrally uncorrelated biphotons can be used to prepare pure single-photons and entangled photons which can be applied for sensing, imaging, and communication with a quantum-enhanced performance at MIR region.
As shown in the  section of Calculation and simulation, all the poling periods of the doped PPLN crystals  are above 5 $\mu$m, which can be easily fabricated using the off-the-shelf technology.
So, the MIR-band single-photon source calculated in the work is ready for fundamental research and industry applications, so as to promote the second quantum revolution.

\section{Conclusion}
In conclusion, we have theoretically studied the preparation of spectrally uncorrelated biphotons at MIR wavelengths from doped LN crystals, including MgO doped LN, ZnO doped LN,  and In$_2$O$_3$ doped ZnLN with doping ratio from 0 to 7 mol\%.
The tilt angle, poling period, thermal properties, and HOM interference of the biphotons are calculated under type-II, type-I, and type-0 phase-matching conditions.
For 5 mol\% MgO doped LN, the three GVM wavelengths are 2474.7 nm, 4016.6 nm, and 3207.6 nm, and the corresponding  purities are 0.968, 0.963, and 0.826, respectively.
In the calculation of the thermal properties, we found that thermal-related GVM wavelength is not dominated by the doped elements, but by   LiNbO$_3$. When the temperature was increased from 20 $^{\circ}C$ to 120 $^{\circ}C$, the three GVM wavelengths are increased by  about 2.8 nm, -78 nm, and -40 nm, respectively.
In the simulation of HOM interference using the  5 mol\% MgO doped PPLN,  visibility of 96.8\% was achieved in a HOM interference between two independent sources, and   visibility of 100\% was achieved in a HOM interference between the signal and idler from the same SPDC source.
The interference patterns oscillate by changing the doping ratio or temperature.
The spectrally uncorrelated biphotons can be used to prepare pure single-photon source and entangled photon source at MIR wavelengths.

\section*{Acknowledgments}
 This work is  partially supported by the National Key R$\&$D Program of China (Grant No. 2018YFA0307400), the National Natural Science Foundations of China (Grant Nos. 91836102, 11704290, 12074299, 61775025, 61405030),  and  the Program of State Key Laboratory of Quantum Optics and Quantum Optics Devices (No: KF201813).
 We thank Profs. Zhi-Yuan Zhou, Yan-Xiao Gong, and Ping Xu for helpful discussions.

\section*{Appendix : The effective nonlinear coefficient}
\begin{figure}[!tbp]
\centering\includegraphics[width=8.5cm]{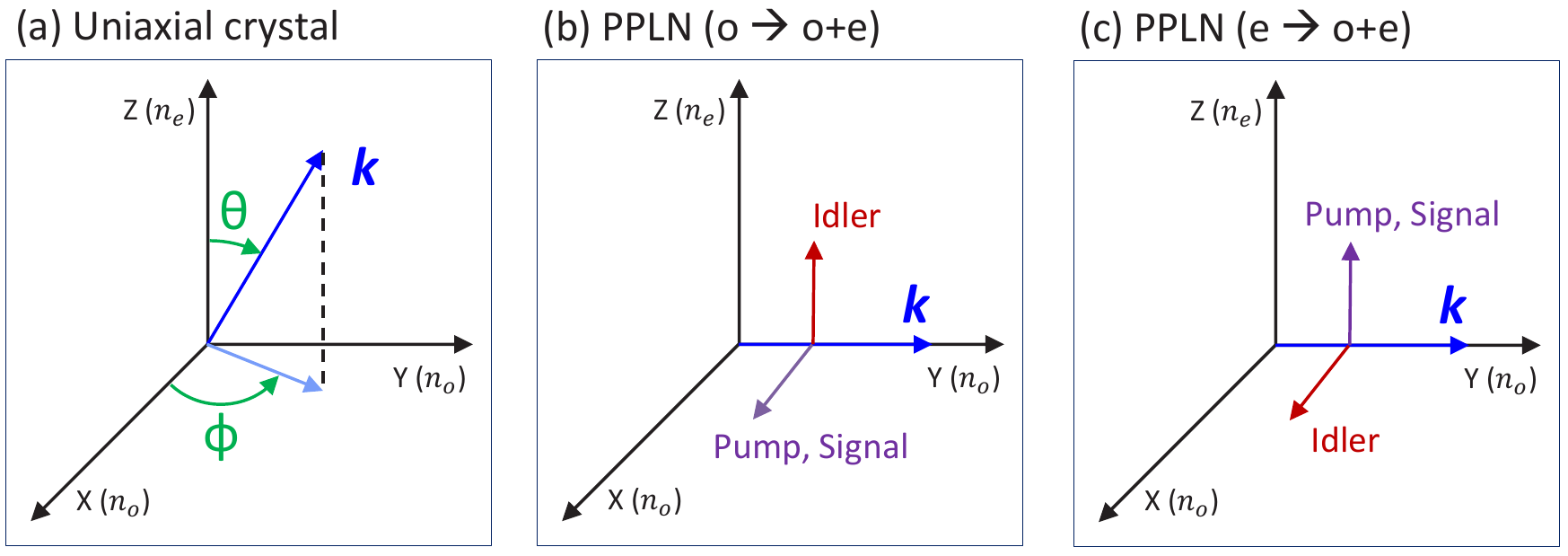}
\caption{ (a) The general refractive index coordinates for uniaxial crystals. (b, c) The refractive index coordinate for PPLN crystal under ($o \to o + e$) or ($e \to o + e$) type-II phase-matching condition.
 } \label{figA1}
\end{figure}

The effective nonlinear coefficient $d_\textrm{eff}$ for collinear phase-matching in uniaxial crystals can be deduced as follow \cite{Smith2016Book}.\\
For $o \to o + e$ phase-matching condition,
\begin{equation}
\begin{array}{l}
d_\textrm{eff}\textrm{(ooe)}= \\
-d_{xxx}C_\theta C_\phi S_{\phi}^2+d_{xyy}C_\theta C_\phi S_{\phi}^2-d_{xyz}S_\theta C_\phi S_\phi \\
+   d_{xxz}S_\theta S_{\phi}^2+d_{xxy}(C_\theta C_{\phi}^2 S_\phi- C_\theta S_{\phi}^3) \\
+d_{yxx}C_\theta C_{\phi}^2 S_\phi -  d_{yyy}C_\theta C_{\phi}^2 S_\phi+d_{yyz}S_\theta C_{\phi}^2 \\
-d_{yxz}S_\theta C_\phi S_\phi +d_{yxy}(C_\theta C_\phi S_{\phi}^2- C_\theta C_{\phi}^3).\\
\end{array}
\end{equation}
For $e \to o + e$ phase-matching condition,
\begin{equation}
\begin{array}{l}
d_\textrm{eff}\textrm{(eoe)}= \\
d_{xxx}C_\theta^2 C_\phi^2 S_\phi-d_{xyy}C_\theta^2 C_\phi^2 S_\phi+d_{xyz}C_\theta S_\theta C_\phi^2 \\
-d_{xxz}C_\theta S_\theta C_\phi S_\phi+d_{xxy}(C_\theta^2 C_\phi S_\phi^2- C_\theta^2 C_\phi^3)\\
+d_{yxx}C_\theta^2 C_\phi S_\phi^2-d_{yyy}C_\theta^2 C_\phi S_\phi^2+d_{yyz}C_\theta S_\theta C_\phi S_\phi \\
-d_{yxz}C_\theta S_\theta S_\phi^2+d_{yxy}(C_\theta^2 S_{\phi}^3- C_\theta^2 C_{\phi}^2 S_\phi)\\
-d_{zxx}C_\theta S_\theta C_\phi S_\phi+d_{zyy}C_\theta S_\theta C_\phi S_\phi-d_{zyz}S_\theta^2 C_\phi\\
+d_{zxz}S_\theta^2 S_\phi+d_{zxy}(C_\theta S_\theta C_\phi^2-C_\theta S_\theta S_\phi^2),\\
\end{array}
\end{equation}
where  $S_\theta=\sin(\theta +\rho)$, $C_\theta=\cos(\theta +\rho)$, $S_\phi=\sin\phi$, $C_\phi=\cos\phi$, and  $\rho$ is the walk-off angle.
$\theta$ is the polar angle, $\phi$ is the azimuth angle, and Z is the optical axis, as defined in  Fig.\,\ref{figA1}(a).
For PPLN under $o \to o + e$ or $e \to o + e$   phase-matching conditions, $\rho=0^\circ$, $\theta=90^\circ$,   and $\phi=90^\circ$, as shown in  Fig.\,\ref{figA1}(b, c).
Therefore,
\begin{equation}
d_\textrm{eff}\textrm{(ooe)}=d_{xxz}
\end{equation}
and
\begin{equation}
d_\textrm{eff}\textrm{(eoe)}=d_{zxz}.
\end{equation}

The second-order nonlinear coefficient matrix for LN is \cite{SNLO70}:
\begin{equation}\label{eff}
\begin{array}{ll}
d&=\left(
  \begin{array}{cccccc}
    0 & 0 &0 & 0 & d_{xxz} & d_{xxy} \\
    d_{yxx} & d_{yyy} & 0 & d_{yyz} & 0 & 0 \\
    d_{zxx} & d_{zyy} & d_{zzz} & 0 & 0 & 0 \\
  \end{array}
\right) \\\\
&=\left(
  \begin{array}{cccccc}
    0 & 0 &0 & 0 & d_{15} & d_{16}\\
   d_{21} & d_{22} & 0 & d_{24} & 0 & 0 \\
    d_{31} & d_{32} &d_{33} & 0 & 0 & 0 \\
  \end{array}
\right) \\ \\
&=\left(
  \begin{array}{cccccc}
    0 & 0 &0 & 0 & -4.6 & -2.2 \\
    -2.2 & 2.2 & 0 & -4.6 & 0 & 0 \\
    -4.6 & -4.6 & -25.0 & 0 & 0 & 0 \\
  \end{array}
\right). \\
\end{array}
\end{equation}

So,
\begin{equation}
d_\textrm{eff}\textrm{(ooe)}=d_{xxz}=-4.6 \, \textrm{pm/V}
\end{equation}
and
\begin{equation}
d_\textrm{eff}\textrm{(eoe)}=d_{zxz}=0.
\end{equation}
Therefore, the configuration of  $e \to o + e$ does not exist for PPLN.

%
%
%

%


%
%
%

%

\end{document}